\documentclass[10pt,journal]{IEEEtran}
\usepackage{scalerel}
\usepackage{rotating}
\usepackage{amssymb}
\usepackage{amsmath}

\usepackage{algorithm, algorithmicx, algpseudocode}
\usepackage{pdfpages}
\usepackage{url}
\usepackage{lineno}
\usepackage{enumitem}
\usepackage{mathtools}
\usepackage{graphicx}
\usepackage{booktabs}

\usepackage{array,multirow}
\usepackage{amsthm}
\usepackage{threeparttable}
\usepackage{mathtools}
\usepackage{booktabs}
\usepackage{amsmath}
\usepackage{geometry}
\usepackage{longtable}
\usepackage{pdflscape}
\usepackage{tabularx}
\usepackage{multicol}
\usepackage{float}
\usepackage{adjustbox}
\usepackage{tablefootnote}
\usepackage{lipsum}
\usepackage{algorithmicx}
\newtheorem{theorem}{Theorem}

\usepackage{tikz}
\usetikzlibrary{shapes.geometric, arrows}
\usetikzlibrary{matrix, positioning}
\usetikzlibrary{fit} 
\usetikzlibrary{calc}

\tikzstyle{decision} = [diamond, text centered, draw=black]
\usepackage{pgfplots}
\pgfplotsset{compat=1.8} 
\usepackage{pgfplotstable}
\usepackage{placeins}
\usetikzlibrary{patterns}
\usepackage{pgfplotstable}
\usepgfplotslibrary{statistics}
\usepackage{tikz}
\usetikzlibrary{shapes.geometric, arrows.meta, positioning}

\tikzstyle{startstop} = [rectangle, rounded corners, minimum width=3cm, minimum height=1cm,text centered, draw=black]
\tikzstyle{process} = [rectangle, minimum width=3cm, minimum height=1cm, text centered, draw=black]
\tikzstyle{decision} = [diamond, minimum width=3cm, minimum height=1cm, text centered, draw=black]
\tikzstyle{arrow} = [thick,->,>=stealth]
\usepackage{pst-platon}
\usetikzlibrary{folding}
\usepackage{multirow}
\usepackage{makecell}
\usepackage{xcolor, soul}
\sethlcolor{yellow}
\usepackage[colorlinks,citecolor=blue,linkcolor=blue]{hyperref}
\makeatletter
\pretocmd{\NAT@open}{\begingroup\color{\@citecolor}}{}{}
\apptocmd{\NAT@close}{\endgroup}{}{}
\makeatother
\usepackage{xstring}
\makeatletter
\AtBeginDocument{
\let\oldref\ref
\renewcommand{\ref}[1]{\IfBeginWith{#1}{fig:}%
{{\color{blue}Fig.~\oldref{#1}}}%
{\IfBeginWith{#1}{lem:}{{\color{blue}Lemma~\oldref{#1}}}
{\IfBeginWith{#1}{tab:}{{\color{blue}Table~\oldref{#1}}}
{\IfBeginWith{#1}{sec:}{{\color{blue}Section~\oldref{#1}}}
{\IfBeginWith{#1}{eq:}{{\color{blue}Eq.~(\oldref{#1})}}
{\IfBeginWith{#1}{ex:}{{\color{blue}Example~(\oldref{#1})}}
{\IfBeginWith{#1}{alg:}{{\color{blue}Algorithm~\oldref{#1}}}
~}}}}%
}}}}
\makeatother
\usepackage{cleveref}
\usepackage{stfloats}
\usepackage[english]{babel}
\makeatletter
\def\ps@pprintTitle{%
  \let\@oddhead\@empty
  \let\@evenhead\@empty
  \let\@oddfoot\@empty
  \let\@evenfoot\@oddfoot
}
\makeatother

\date{}
\usepackage{orcidlink}
\begin{document}
\title{Optimizing Scalar Selection in Elliptic Curve Cryptography Using Differential Evolution  for Enhanced Security}
\author{Takreem Haider\,\orcidlink{0000-0002-9224-4576}%
\thanks{Takreem Haider is with the Department of Basic Sciences $\&$ Humanities, College of Electrical and Mechanical Engineering, National University of Sciences and Technology, Islamabad, Pakistan (email:takreem.haider@ceme.nust.edu.pk).} }

\maketitle

\begin{abstract}
Elliptic Curve Cryptography (ECC) is a fundamental component of modern public-key cryptosystems that enable efficient and secure digital signatures, key exchanges, and encryption. Its core operation, scalar multiplication, denoted as $k \cdot P$, where $P$ is a base point and $k$ is a private scalar, relies heavily on the secrecy and unpredictability of $k$. Conventionally, $k$ is selected using user input or pseudorandom number generators. However, in resource-constrained environments with weak entropy sources, these approaches may yield low-entropy or biased scalars, increasing susceptibility to side-channel and key recovery attacks. To mitigate these vulnerabilities, we introduce an optimization-driven scalar generation method that explicitly maximizes bit-level entropy. Our approach uses differential evolution (DE), a population-based metaheuristic algorithm, to search for scalars whose binary representations exhibit maximal entropy, defined by an even and statistically uniform distribution of ones and zeros. This reformulation of scalar selection as an entropy-optimization problem enhances resistance to entropy-based cryptanalytic techniques and improves overall unpredictability. Experimental results demonstrate that DE-optimized scalars achieve entropy significantly higher than conventionally generated scalars.  The proposed method can be integrated into existing ECC-based protocols, offering a deterministic, tunable alternative to traditional randomness, ideal for applications in blockchain, secure messaging, IoT, and other resource-constrained environments.
\end{abstract}


\section{Introduction}
\IEEEPARstart{E}LLIPTIC curve cryptography (ECC) has gained significant attention due to its ability to offer security comparable to that of classical public-key cryptosystems, such as Rivest–Shamir–Adleman~(RSA), while requiring substantially shorter key lengths. For instance, a 163-bit ECC key provides a security level equivalent to a 1024-bit RSA key~\cite{koblitz2000state}. This high level of efficiency makes ECC particularly suitable for implementation in resource-constrained environments. In addition, ECC supports critical cryptographic functions, including digital signatures, secure key exchange, and end-to-end encryption, particularly in settings such as mobile devices, embedded systems, and Internet of Things (IoT) networks~\cite{miller1985use, koblitz1987elliptic}.

At the heart of the ECC lies the scalar point multiplication operation, denoted as $k \cdot P$, where $P$ is a publicly known base point on the elliptic curve (EC) and $k$ is a large secret scalar chosen from a finite field. This operation forms the cryptographic backbone of widely used protocols such as the elliptic curve digital signature algorithm (ECDSA) and elliptic curve Diffie–Hellman (ECDH). The security of these algorithms is fundamentally based on the hardness of the elliptic curve discrete logarithm problem (ECDLP), which assumes that computing $k$ from the pair $(P, k \cdot P)$ is computationally infeasible.

Although substantial research has focused on optimizing the arithmetic and propagation time of scalar multiplication, comparatively little attention has been paid to the entropy and unpredictability of the scalar $k$ itself. In existing EC-based systems, $k$ is user-specified or generated using pseudorandom number generator (PRNG). However, such methods are not always reliable, particularly in constrained environments where entropy sources are weak or poorly implemented. Low-entropy or biased scalar values can leak information through side-channel emissions or enable cryptanalytic key-recovery attacks, thus undermining the intended security guarantees of ECC systems.

To address this, we propose a novel optimization-based framework for secure scalar generation. Our method treats scalar selection as a bit-level entropy maximization problem and uses differential evolution (DE), a robust population-based metaheuristic to search for scalar values whose binary representations demonstrate high entropy and uniform bit distribution. By explicitly targeting statistical unpredictability in scalar generation, the proposed approach enhances the resistance of ECC implementations to side-channel, key-recovery, and entropy-based attacks.

The main contributions of this work are as follows.
\begin{itemize}
    \item We introduce an optimization-driven approach for scalar generation in ECC by maximizing binary entropy to strengthen the statistical resilience of the private key.
    \item We use DE, a population-based metaheuristic algorithm, to explore the scalar space, demonstrating its superiority over traditional pseudorandom scalar selection, especially under low-entropy or adversarial conditions.
    \item We conducted extensive experiments that compare entropy, bit uniformity, and unpredictability metrics between DE-optimized scalars and conventional random scalars, showing significant improvements in security-critical dimensions.
\end{itemize}

The remainder of the paper is organized as follows. 
\ref{sec:related_work} provides an overview of related work in ECC, emphasizing its role in the construction of secure cryptographic primitives such as PRNGs, substitution boxes (S-boxes), encryption schemes, and digital watermarking techniques.~\ref{sec:preliminaries} presents a relevant background on finite field arithmetic, EC fundamentals, and scalar multiplication algorithms.~\ref{sec:scheme} details the proposed entropy-optimized scalar generation method.~\ref{sec:analysis} discusses the experimental setup, evaluation metrics, and comparative results. Finally,~\ref{sec:conclusion} provides concluding remarks and outlines the directions for future work.
\section{Literature Review}\label{sec:related_work}
The challenges of optimizing the cryptographic systems of ECs are essential to advance their security, computational efficiency, and real-world applicability. The pioneering works of Miller~\cite{miller1985use} and Koblitz~\cite{koblitz1987elliptic} introduced the use of ECs in public-key cryptography, highlighting the importance of judicious parameter selection to achieve strong cryptographic guarantees. Washington~\cite{washington2008elliptic} further emphasized the influence of critical parameters such as curve coefficients, base point, field size, and key lengths on the performance and security of the ECC.
Lenstra and Verheul~\cite{lenstra2001selecting} reinforced the need to select robust ECC parameters, particularly focusing on the choice of the prime modulus. Blake et al.~\cite{blake1999elliptic} provided a comprehensive overview of EC parameterization, discussing strategies to select optimal base points, curve structures, and finite fields to balance efficiency with cryptographic strength. Moreover, Hankerson et al.~\cite{hankerson2004guide} addressed performance concerns by presenting a variety of algorithmic techniques to accelerate point multiplication, a core operation in ECC.

In recent developments, ECs have found widespread applications in the design of modern cryptographic primitives,  PRNGs~\cite{haider2024novel, liu2025constructing}, S-boxes design~\cite{haider2024substitution, ali2024novel}, image encryption~\cite{haider2023novel, sattar2023efficient}, and digital watermarking~\cite{azam2024optimized, zhang2024provably}. 
For instance, Haider et al.~\cite{haider2024novel} proposed an image-dependent PRNG that integrates ECs with a multi-objective genetic algorithm, offering strong randomness characteristics. In addition, Haider et al.~\cite{haider2024substitution} developed an efficient S-box generation framework comprising three main stages, beginning with the generation of EC points as seeds for substitution mapping. Similarly, Azam et al.~\cite{azam2024optimized} introduced a watermarking technique that utilizes EC-based pseudorandom sequences and genetic algorithms to determine embedding positions with enhanced unpredictability and robustness.

Despite significant progress in ECC-based cryptographic schemes, the selection and optimization of the scalar \(k\), a fundamental parameter used in ECC operations, remains largely underexplored. 
Metaheuristic algorithms such as genetic algorithms, particle swarm optimization, and simulated annealing have been successfully applied to various cryptographic problems, including key scheduling, S-box construction, and parameter tuning~\cite{singh2025optimizing, haider2024substitution, mullai2021enhancing}. However, their application to scalar optimization in EC systems is minimal.  DE, introduced by Lampinen et al.~\cite{lampinen2005differential}, is a well-established method for solving nonlinear optimization problems and has shown effectiveness in domains such as biometric protection~\cite{ng2024palmprint}, encryption~\cite{toktas2021chaotic}, and cryptographic key generation~\cite{krishna2018key}.

To the best of our knowledge, the present work is the first to apply an evolutionary optimization technique specifically aimed at maximizing the bit-level entropy of ECC scalars. By framing scalar selection as an entropy-driven optimization problem, our method ensures balanced bit distributions, enhancing resistance to statistical and side-channel attacks. This approach offers a tunable and deterministic alternative to traditional randomness sources, positioning our contribution at the intersection of evolutionary computation and cryptographic key generation.
\section{Background on Elliptic Curves}\label{sec:preliminaries}

Let $\mathbb{F}_q$ be a finite field with a characteristic greater than 3. An EC $E$ defined over $\mathbb{F}_q$ can be expressed by the Weierstrass equation~\cite{hankerson2004guide}:
\begin{equation}
    y^2 = x^3 + ax + b,
\end{equation}
where $a, b \in \mathbb{F}_q$, and the condition $4a^3 + 27b^2 \ne 0$ ensures that the curve is non-singular. The set of points $(x, y) \in \mathbb{F}_q \times \mathbb{F}_q$ satisfying this equation, along with a distinguished point at infinity (often denoted  $\mathcal{O}$), constitutes the curve $E(\mathbb{F}_q)$.
The set $E(\mathbb{F}_q)$ forms an abelian group under point addition, with $\mathcal{O}$ serving as the identity element. For any two points $P = (x_1, y_1)$ and $Q = (x_2, y_2)$ on the curve, the sum $R = P + Q = (x_3, y_3)$ is computed as follows~\cite{hankerson2004guide}:
\begin{align}
    \lambda &=
    \begin{cases}
        \dfrac{y_2 - y_1}{x_2 - x_1} & \text{if } P \ne Q, \\
        \dfrac{3x_1^2 + a}{2y_1} & \text{if } P = Q,
    \end{cases} \\
    x_3 &= \lambda^2 - x_1 - x_2, \\
    y_3 &= \lambda(x_1 - x_3) - y_1.
\end{align}
In the case of fields with characteristic 2, ECs are represented using alternative forms. For instance, a curve with zero $j$-invariant can be written as:
\begin{equation}\label{eq:EC_1}
    y^2 + cy = x^3 + ax + b,
\end{equation}
where $a, b, c \in \mathbb{F}_q$ and $c \ne 0$. Another common form, for curves with nonzero $j$-invariant, is:
\begin{equation}\label{eq:EC_2}
    y^2 + xy = x^3 + ax^2 + b,
\end{equation}
with $a, b \in \mathbb{F}_q$ and $b \ne 0$. In both cases, the point at infinity serves as the group identity.

For the curve in~\ref{eq:EC_1}, the point negation is given by $-P = (x_1, y_1 + c)$ for $P = (x_1, y_1)$. If $P \ne Q$, then the addition $P + Q = (x_3, y_3)$ is defined by~\cite{hankerson2004guide}:

\begin{align}
    x_3 &=
    \left(\frac{y_1 + y_2}{x_1 + x_2}\right)^2 + x_1 + x_2, \\
    y_3 &=
    \left(\frac{y_1 + y_2}{x_1 + x_2}\right)(x_1 + x_3) + y_1 + c.
\end{align}

For the special case $P = Q$, the formulas are as follows:

\begin{align}
    x_3 &= \left(\frac{x_1^2 + a}{c}\right)^{2}, \\
    y_3 &= \left(\frac{x_1^2 + a}{c}\right)(x_1 + x_3) + y_1 + c.
\end{align}

For curves in~\ref{eq:EC_2}, the negation rule is $-P = (x_1, y_1 + x_1)$. Addition and doubling are defined similarly, but with slight modifications:

\begin{align}
    x_3 &=
    \begin{cases}
        \left(\frac{y_1 + y_2}{x_1 + x_2}\right)^2 + \frac{y_1 + y_2}{x_1 + x_2} + x_1 + x_2 + a & \text{if } P \ne Q, \\
        x_1^2 + \frac{b}{x_1^2} & \text{if } P = Q,
    \end{cases} \\
    y_3 &=
    \begin{cases}
        \left(\frac{y_1 + y_2}{x_1 + x_2}\right)(x_1 + x_3) + x_3 + y_1 & \text{if } P \ne Q, \\
        x_1^2 + \left(\frac{x_1 + y_1}{x_1}\right)x_3 + x_3 & \text{if } P = Q.
    \end{cases}
\end{align}

\begin{theorem}[Hasse's Theorem~\cite{hankerson2004guide}]
Let $E$ be an EC defined over a finite field $\mathbb{F}_q$. Then the total number of $\mathbb{F}_q$-rational points on $E$, denoted $\#E(\mathbb{F}_q)$, satisfies the inequality:
\[
\left| \#E(\mathbb{F}_q) - (q + 1) \right| \leq 2\sqrt{q}.
\]
\end{theorem}

\section{Proposed Methodology}\label{sec:scheme}
This section provides a comprehensive overview of the proposed methodology and its underlying principles.
The security of ECC is fundamentally based on the secrecy and randomness of the scalar used in the point multiplication operation. Traditional scalar selection mechanisms, often based on PRNGs, can be vulnerable in constrained environments due to insufficient entropy or biased distributions. To mitigate this issue, we propose a novel optimization-based scalar generation framework that uses DE, a population-based metaheuristic algorithm, to generate scalars with maximal bit-level entropy.
\subsection{Phase 1: Selection of Elliptic Curve}
We begin by selecting a secure EC defined over a finite field $\mathbb{F}_q$. The EC is given by the short Weierstrass form:
\begin{equation}
E: y^2 \equiv x^3 + ax + b,
\end{equation}
where $a, b \in \mathbb{F}_q$ and the discriminant condition $4a^3 + 27b^2 \neq 0 $ ensures that the curve is non-singular. Let $\mathcal{P}$ denote the set of $\mathbb{F}_q$-rational points on $E$, including the point at infinity $\mathcal{O}$. The order of the curve, denoted by $n = \#E(\mathbb{F}_q)$, is typically a large number.
In EC cryptosystems, the primary cryptographic operation is scalar multiplication.
\begin{equation}
Q = k \cdot P,
\end{equation}
where $P \in E(\mathbb{F}_q)$ is a publicly known base point of order $n$, $k \in \mathbb{Z}_n$ is a secret scalar and $Q$ is the resulting point.

\subsection{Phase 2: Entropy-Based Formulation of Scalar Selection}
To enhance the cryptographic strength of scalar selection, we reformulate the problem of generating $k$ as an entropy-maximization task. Specifically, we define an objective function $\mathcal{H}(k)$ that computes the Shannon entropy of the binary representation of $k$:
\begin{equation}
\mathcal{H}(k) = -p_0 \log_2(p_0) - p_1 \log_2(p_1),
\end{equation}
where $p_0$ and $p_1$ represent the proportions of zeros and ones in the binary representation of $k$, respectively. For maximum entropy, we aim for $p_0 \approx p_1 \approx 0.5$.

\subsection{Phase 4: Application of Differential Evolution}
To find a scalar $k$ that maximizes $\mathcal{H}(k)$, we utilize DE, a population-based optimization algorithm known for its efficiency and robustness.

\subsubsection*{Step 1: Initialization}
A population of $M$ candidate scalars $\{k_i\}_{i=1}^{M}$ is randomly initialized within the interval $[1, n-1]$, ensuring that each candidate is a valid scalar.
\begin{equation}
k_i \in \mathbb{Z}_n, \quad \forall i \in \{1, 2, ..., M\}.
\end{equation}

\subsubsection*{Step 2: Mutation}
For each candidate scalar $k_i$, a mutant vector $v_i$ is generated using three distinct individuals $k_{r1}$, $k_{r2}$, and $k_{r3}$ from the population:
\begin{equation}
v_i = k_{r1} + m_r \cdot (k_{r2} - k_{r3}) \pmod{n}, 
\end{equation}
where $m_r \in (0,1)$ is the differential weight that controls the amplification of the differential variation.

\subsubsection*{Step 3: Crossover}
A trial vector $u_i$ is generated by recombining the original vector $k_i$ and its mutant $v_i$:
\begin{equation}
u_{i,j} =
\begin{cases}
v_{i,j} & \text{if } \text{rand}_j \leq c_r \text{ or } j = j_{\text{rand}}, \\
k_{i,j} & \text{otherwise}.
\end{cases}
\end{equation}
Here, $c_r \in [0,1]$ is the crossover rate and $j_{\text{rand}}$ ensures that at least one component is inherited from the mutant vector.

\subsubsection*{Step 4: Selection}
The new scalar $k_i$ is updated based on a comparison of entropy values.
\begin{equation}
k_i^{(t+1)} =
\begin{cases}
u_i & \text{if } \mathcal{H}(u_i) > \mathcal{H}(k_i), \\
k_i & \text{otherwise}.
\end{cases}
\end{equation}

\subsubsection*{Step 5: Termination}
The DE algorithm continues until a termination criterion is met, such as a maximum number of generations or convergence to a high-entropy scalar. The final output $k_{\text{optimized}}$ is the scalar with the highest entropy:
\begin{equation}
k_{\text{opt}} = \arg\max_{k_i \in \text{population}} \mathcal{H}(k_i).
\end{equation}

\subsection{Phase 5: Integration with ECC}
Once the optimized scalar $k_{\text{opt}}$ is selected, it can be used directly in scalar multiplication to produce the public point.
\begin{equation}
Q = k_{\text{optimized}} \cdot P.
\end{equation}
This approach ensures that the scalar used in ECC computations is statistically robust, resistant to entropy-based cryptanalytic attacks, and suitable for integration into existing protocols such as ECDSA and ECDH without modifying the core ECC operations.

\ref{fig:flowchart} illustrates the overall flow diagram of the proposed scheme, providing an overview of its operational steps. \ref{alg:DE_entropy} presents the corresponding pseudocode, detailing the step-by-step procedure of the method. To further clarify the scheme, Example 1 demonstrates each step using a small finite field, offering a concrete and illustrative example of the proposed approach.

By treating scalar selection as an optimization problem centered on entropy maximization, the proposed methodology provides a deterministic yet highly secure alternative to PRNG-based scalar generation. The use of DE allows for a tunable trade-off between computational cost and security, making the method applicable in both high-performance and resource-constrained cryptographic environments.

\begin{algorithm}[t!]
\caption{Elliptic curve scalar optimization using differential evolution algorithm.}
\label{alg:DE_entropy}
\begin{algorithmic}[1]
\Require EC $E: y^2 = x^3 + ax + b$, base point $P \in E(\mathbb{F}_q)$ of order $n$, DE parameters: population size $M$, mutation factor $m_r$, crossover rate $c_r$, max generations $\ell$.
\Ensure Optimized scalar $k_{\text{opt}}$ with maximum entropy.
\State \textbf{Initialization:}
\State Generate initial population $\mathcal{K}^{(0)} = \{k_1, k_2, ..., k_M\}$ randomly from $[1, n - 1]$
\For{generation $t = 1$ to $\ell$}
    \For{each individual $k_i$ in $\mathcal{K}^{(t)}$}
        \State Randomly choose distinct indices $r_1, r_2, r_3\in \{1, ..., M\}$ such that $r_1 \ne r_2 \ne r_3 \ne i$
        \State \textbf{Mutation:} $v_i \gets (k_{r_1} + m_r \cdot (k_{r_2} - k_{r_3})) \pmod{n}$
        \State \textbf{Crossover:}
        \State Initialize trial vector $u_i$
        \For{each bit position $j$ in $k_i$}
            \If{rand() $\le$ $c_r$ or $j = j_{\text{rand}}$}
                \State $u_{i,j} \gets v_{i,j}$
            \Else
                \State $u_{i,j} \gets k_{i,j}$
            \EndIf
        \EndFor
        \State \textbf{Selection:}
        \State Compute entropy $\mathcal{H}(k_i)$ and $\mathcal{H}(u_i)$:
        \[
            \mathcal{H}(k) = -p_0 \log_2(p_0) - p_1 \log_2(p_1)
        \]
        \State where $p_0$ and $p_1$ are the probabilities of 0 and 1 bits in $k$
        \If{$\mathcal{H}(u_i) > \mathcal{H}(k_i)$}
            \State $k_i^{(t+1)} \gets u_i$
        \Else
            \State $k_i^{(t+1)} \gets k_i$
        \EndIf
    \EndFor
\EndFor
\State \textbf{Output:} $k_{\text{opt}} \gets \arg\max_{k_i \in \mathcal{K}^{(G)}} \mathcal{H}(k_i)$
\end{algorithmic}
\end{algorithm}
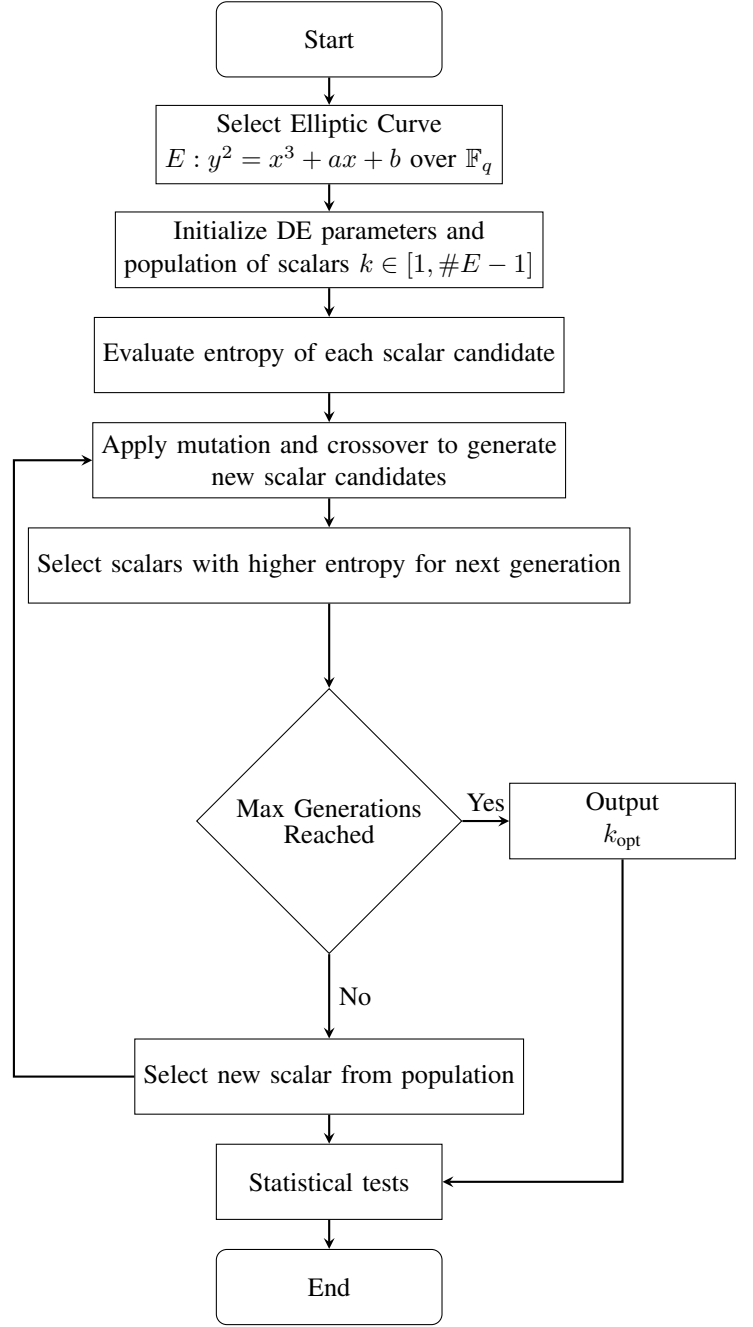
\begin{figure}[t!]
\centering
\begin{tikzpicture}[node distance=1.4cm]

\node (start) [startstop] {Start};
\node (selectEC) [process, below of=start] 
  {\shortstack{Select Elliptic Curve\\ $E: y^2 = x^3 + ax + b $ over $\mathbb{F}_q$ }};
\node (initDE) [process, below of=selectEC]  
  {\shortstack{Initialize DE parameters and \\ population of scalars $k \in [1, \#E -1]$}};
\node (evaluate) [process, below of=initDE] 
  {Evaluate entropy of each scalar candidate};
\node (mutate) [process, below of=evaluate] 
  {\shortstack{Apply mutation and crossover to generate\\ new scalar candidates}};
\node (select) [process, below of=mutate] 
  {Select scalars with higher entropy for next generation};
\node (checkGen) [decision, below of=select, yshift=-2.0cm]
  {\shortstack{Max Generations \\Reached}};
\node (output) [process, right of=checkGen, xshift=2.5cm] 
  {\shortstack{Output\\ $k_{\text{opt}}$}};
\node (computeKP) [process, below of=checkGen, yshift=-2.0cm] 
  {Select new scalar from population};
\node (validate) [process, below of=computeKP] 
  {Statistical tests};
\node (end) [startstop, below of=validate] {End};

\draw [arrow] (start) -- (selectEC);
\draw [arrow] (selectEC) -- (initDE);
\draw [arrow] (initDE) -- (evaluate);
\draw [arrow] (evaluate) -- (mutate);
\draw [arrow] (mutate) -- (select);
\draw [arrow] (select) -- (checkGen);
\draw [arrow] (checkGen) -- node[anchor=south] {Yes} (output);
\draw [arrow] (checkGen) -- node[anchor=west] {No} (computeKP);
\draw [arrow] (output) |- (validate);
\draw [arrow] (computeKP) -- (validate);
\draw [arrow] (validate) -- (end);

\draw [arrow] (computeKP.west) -- ++(-1.6,0) |- (mutate.west);
\end{tikzpicture}
\caption{Flowchart of the proposed entropy-maximized scalar generation scheme using differential evolution.}
\label{fig:flowchart}
\end{figure}
\bigskip
\noindent
{\bf{Example 1}}
[EC over $\mathbb{F}_{29}$ with DE-Based Scalar Selection]
Let the finite field be $\mathbb{F}_{29}$, and define the EC 
\(
y^2 = x^3 + 4x + 20 \pmod{29},
\)
with coefficients \( a = 4 \), \( b = 20 \), and prime \( p = 29 \).
To verify non-singularity, we compute the discriminant:
\(
\Delta = -16(4a^3 + 27b^2) = -16(4 \cdot 64 + 27 \cdot 400) = -16(256 + 10800) = -16 \cdot 11056.
\)
Reducing modulo 29:
\(
\Delta \equiv -176896 \equiv 8 \pmod{29} \neq 0,
\)
which confirms that the curve is non-singular over $\mathbb{F}_{29}$.

\noindent
\textit{Rational Points on \( E(\mathbb{F}_{29}) \):}  
There are 36 affine points and the point at infinity \(\mathcal{O}\), giving:
\(
\#E(\mathbb{F}_{29}) = 37.
\)
\begin{center}
\begin{tabular}{|c|c|c|c|}
\hline
$(0,7)$   & $(0,22)$  & $(1,5)$   & $(1,24)$ \\
$(2,6)$   & $(2,23)$  & $(3,1)$   & $(3,28)$ \\
$(4,10)$  & $(4,19)$  & $(5,7)$   & $(5,22)$ \\
$(6,12)$  & $(6,17)$  & $(8,10)$  & $(8,19)$ \\
$(10,4)$  & $(10,25)$ & $(13,6)$  & $(13,23)$ \\
$(14,6)$  & $(14,23)$ & $(15,2)$  & $(15,27)$ \\
$(16,2)$  & $(16,27)$ & $(17,10)$ & $(17,19)$ \\
$(19,13)$ & $(19,16)$ & $(20,3)$  & $(20,26)$ \\
$(24,7)$  & $(24,22)$ & $(27,2)$  & $(27,27)$ \\
\hline
\end{tabular}
\end{center}
\noindent
\textit{Base Point:} Let \( P = (0, 7) \in E(\mathbb{F}_{29}) \).  
We apply the DE algorithm to select a scalar \( k \in [1, 36] \) that maximizes the entropy of its binary representation.  
DE is configured with:
\(
\text{population size} =M=20, \quad \text{crossover rate} =c_r=0.9, \quad \text{mutation factor} =m_r=0.8, \quad \text{total number of generations} = \ell = 50.
\)

\noindent
\textit{Result:} After optimization, assume the DE algorithm selects:
\(
k_{\text{opt}} = 19, \quad \text{so} \quad k_{\text{opt}} \cdot P = (19, 16).
\)
Binary representation:
\(
k_{\text{opt}} = (10011)_2,
\)
with Shannon entropy:
\[
\mathcal{H}(k_{\text{opt}}) = -\sum_{i} p_i \log_2 p_i \approx 0.97095,
\]
where \( p_i \) are the probabilities of 0s and 1s in the bitstring.
This example illustrates the end-to-end process of applying DE for high-entropy scalar generation over ECs, showcasing the integration of evolutionary optimization in cryptographic primitives.

\section{Experimental Results: Analysis and Comparison}\label{sec:analysis}
This section presents the experimental evaluation of the proposed method, focusing on its statistical robustness and cryptographic suitability. Specifically, we benchmark the DE-derived scalar values against scalars selected randomly, using a variety of standard randomness and entropy tests. The goal is to assess whether the optimized scalars exhibit improved characteristics in cryptographic applications.\\

\noindent
\textbf{\textit{Test Environment and System Specifications:}}
All experiments were performed on a personal computer system with the specifications listed in~\ref{tab:system-specs}. This configuration ensures sufficient computational capacity for cryptographic operations and statistical testing.
\begin{table}[ht!]
\centering
\caption{System Specifications}
\label{tab:system-specs}
\begin{tabular}{|l|l|}
\hline
\textbf{Component} & \textbf{Specification} \\
\hline
Processor          & 13th Gen Intel(R) Core(TM) i5-1334U @ 1.30 GHz \\
Installed RAM      & 16.0 GB (15.7 GB usable) \\
System Type        & 64-bit OS, x64-based processor \\
\hline
\end{tabular}
\end{table}

\noindent
\textbf{\textit{Target Cryptographic Curves:}}
Evaluation is carried out using standardized NIST 
ECs~\cite{sp2023recommendations}:
\begin{itemize}
  \item \textbf{NIST P-192}
  \item \textbf{NIST P-224}
  \item \textbf{NIST P-256}
\end{itemize}
These curves are widely recognized in cryptographic standards and serve as a robust foundation for benchmarking scalar behavior. The detailed parameters of these curves are provided in~\ref{tab:nist_curves}.\\
\begin{table*}[t!]
\centering
\caption{Parameters of NIST recommended elliptic curves.}
\begin{tabular}{@{}ll@{}}
\toprule
\textbf{Curve} & \textbf{Parameter Values} \\
\midrule
Curve Name & \textbf{NIST P-192 (secp192r1)}\\
Field Prime $(p)$ & \texttt{0xfffffffffffffffffffffffffffffffeffffffffffffffff} \\
Curve Coefficient $(a)$ & \texttt{0xfffffffffffffffffffffffffffffffefffffffffffffffc} \\
Curve Coefficient $(b)$ & \texttt{0x64210519e59c80e70fa7e9ab72243049feb8deecc146b9b1} \\
Base Point $G = (x, y)$ & 
\begin{tabular}[t]{@{}l@{}}
\texttt{(0x188da80eb03090f67cbf20eb43a18800f4ff0afd82ff1012,} \\
\texttt{\ \ \ 0x07192b95ffc8da78631011ed6b24cdd573f977a11e794811)}
\end{tabular} \\
Order of Base Point  $(n)$ & \texttt{0xffffffffffffffffffffffff99def836146bc9b1b4d22831} \\[1.5ex]
\midrule
Curve Name & \textbf{NIST P-224 (secp224r1)}  \\
Field Prime $(p)$ & \texttt{0xffffffffffffffffffffffffffffffff000000000000000000000001} \\
Curve Coefficient $(a)$ & \texttt{0xfffffffffffffffffffffffffffffffefffffffffffffffffffffffe} \\
Curve Coefficient $(b)$ & \texttt{0xb4050a850c04b3abf54132565044b0b7d7bfd8ba270b39432355ffb4} \\
Base Point $G = (x, y)$ &
\begin{tabular}[t]{@{}l@{}}
\texttt{(0xb70e0cbf6bb4bf7f321390b9 4a03c1d356c21122343280d6115c1d21,} \\
\texttt{\ \ \ 0xbd376388b5f723fb4c22dfe6cd4375a05a07476444d5819985007e34)}
\end{tabular} \\
Order of Base Point $(n)$ & \texttt{0xffffffffffffffffffffffffffff16a2e0b8f03e13dd29455c5c2a3d} \\[1.5ex]
\midrule
Curve Name & \textbf{NIST P-256 (secp256r1)} \\
Field Prime $(p)$ & \texttt{0xffffffff00000001000000000000000000000000ffffffffffffffffffffffff} \\
Curve Coefficient $(a)$ & \texttt{0xffffffff00000001000000000000000000000000fffffffffffffffffffffffc} \\
Curve Coefficient $(b)$ & \texttt{0x5ac635d8aa3a93e7b3ebbd55769886bc651d06b0cc53b0f63bce3c3e27d2604b} \\
Base Point $G = (x, y)$ &
\begin{tabular}[t]{@{}l@{}}
\texttt{(0x6b17d1f2e12c4247f8bce6e563a440f277037d812deb33a0f4a13945d898c296,} \\
\texttt{\ \ \ 0x4fe342e2fe1a7f9b8ee7eb4a7c0f9e162cbf4f3c7e0c8a9b9eebe9e1e6e28238)}
\end{tabular} \\
Order of Base Point $(n)$ & \texttt{0xffffffff00000000ffffffffffffffffbce6faada7179e84f3b9cac2fc632551} \\
\bottomrule
\end{tabular}
\label{tab:nist_curves}
\end{table*}

\noindent
\textbf{\textit{Differential Evolution Parameters:}}
The DE algorithm plays a central role in optimizing scalar values for enhanced randomness. To ensure reproducibility and adherence to commonly accepted practices, the DE was configured using standard parameter values, summarized as follows.
\begin{itemize}
  \item \textbf{Population Size ($M$):} 50 individuals were maintained in each generation. This ensures sufficient genetic diversity while keeping the computational cost manageable.
  \item \textbf{Crossover Rate ($c_r$):} Set to 0.9, promoting high recombination between candidate solutions, which helps explore the search space effectively.
  \item \textbf{Mutation Factor ($m_r$):} Chosen as 0.8 to provide a balance between exploration (diversity) and exploitation (convergence).
  \item \textbf{Maximum Generations ($\ell$):} The evolutionary process was run for up to 100 generations, allowing adequate opportunity for convergence toward optimal or near-optimal solutions.
\end{itemize}

These parameter values were selected based on prior literature and empirical studies, which suggest they offer a good trade-off between optimization quality and computational efficiency in cryptographic parameter tuning.\\

\noindent
\textbf{\textit{Evaluation Metrics and Statistical Tests:}}
To assess the randomness and quality of the scalar values, we used the following standard statistical tests:
\begin{itemize}
  \item \textbf{Shannon Entropy} 
  \item \textbf{Frequency (Monobit) Test} 
  \item \textbf{Chi-Square Test} 
  \item \textbf{Runs Test}
  \item \textbf{Autocorrelation Test}
  \item \textbf{Compression Ratio}
\end{itemize}

The purpose of these experiments is to validate whether the scalars generated by DE optimization demonstrate enhanced statistical properties compared to their randomly selected counterparts. The results serve to reinforce the cryptographic viability of the proposed scalar generation method.
\subsection{Shannon Entropy}
Shannon entropy~\cite{shannon1948mathematical} is a fundamental metric in information theory used to quantify the uncertainty or randomness of a system. For binary sequences composed of bits 0 and 1, the entropy measures the unpredictability of bit values and is defined as:
\begin{equation}
\mathcal{H}(X) = -p_0 \log_2 p_0 - p_1 \log_2 p_1,
\end{equation}
where:
\begin{itemize}
  \item \( p_0 \) is the probability (or proportion) of 0s in the sequence,
  \item \( p_1 \) is the probability (or proportion) of 1s in the sequence,
  \item \( p_0 + p_1 = 1 \).
\end{itemize}
The entropy \( \mathcal{H}(X) \) reaches its maximum value of 1 when \( p_0 = p_1 = 0.5 \), indicating a perfectly random and balanced sequence. Conversely, if all bits in the sequence are the same (i.e., \( p_0 = 1 \) or \( p_1 = 1 \)), the entropy becomes 0, reflecting no uncertainty or randomness.
Higher entropy values indicate better resistance to statistical attacks as a result of increased unpredictability.

We calculated entropy for both the randomly selected scalar \( k \) and the optimized scalar \( k_{\text{opt}} \) on multiple NIST-recommended ECs
The results are presented in~\ref{tab:Entropy_192},~\ref{tab:Entropy_224}, and~\ref{tab:Entropy_256} for the NIST P-192, P-224, and P-256 curves, respectively.
The results showed that the entropy of \( k_{\text{opt}} \) was higher than that of \( k \), often approaching the ideal value of 1. For example, in the case of the NIST P-192 curve, the entropy of the random scalar \( k \) was measured at 0.9949, whereas the optimized scalar \( k_{\text{opt}} \) achieved a value of 1. Similar trends were observed in other tested curves.
These findings support the conclusion that the proposed optimization approach improves the randomness properties of the scalars, a critical factor in ensuring resistance against statistical attacks.
\begin{table*}[t!]
\centering
\caption{Shannon entropy comparison between randomly selected scalar \( k \) and optimized scalar \( k_{\text{opt}} \) using the NIST P-192.}
\begin{tabular}{@{}|l|l|l|l|l|@{}}
\hline
\textbf{Scalar Type} & \textbf{Hex Value}  & \textbf{Shannon Entropy} & \textbf{1's Count} & \textbf{0's Count} \\
\hline
$k$ & \texttt{0x9c6786c6212db513501dd99840e73bb1a2168c652541eb1b} & 0.9949 & 88  & 104 \\
$k_{\text{opt}}$    & \texttt{0xb39b2b88cf9d25687b6b3836aae3f6316b1faef713b505c0} & 1.0000 & 96 & 96 \\
\hline
\end{tabular}
\label{tab:Entropy_192}
\vspace{0.5em}
\caption{Shannon entropy comparison between randomly selected scalar \( k \) and optimized scalar \( k_{\text{opt}} \) using the NIST P-224.}
\begin{tabular}{@{}|l|l|l|l|l|@{}}
\hline
\textbf{Scalar Type} & \textbf{Hex Value}  & \textbf{Shannon Entropy} & \textbf{1's Count} & \textbf{0's Count} \\
\hline
$k$ & \texttt{0xc86d61af1acbfc87f51e62f8c0b93a840d4d984a28442e516004994a} & 0.9917 & 100 & 124 \\
$k_{\text{opt}}$    & \texttt{0xbe4065efd2904203e07be3f64b3d629c8f1d26666f875d1b40f87285} & 1.0000 & 112 & 112 \\
\hline
\end{tabular}
\label{tab:Entropy_224}
\vspace{0.5em}
\caption{Shannon entropy comparison between randomly selected scalar \( k \) and optimized scalar \( k_{\text{opt}} \) using the NIST P-256.}
\setlength{\tabcolsep}{2.5pt}
\begin{tabular}{@{}|l|l|l|l|l|@{}}
\hline
\textbf{Scalar Type} & \textbf{Hex Value}  & \textbf{Shannon Entropy} & \textbf{1's Count} & \textbf{0's Count} \\
\hline
$k$ & \texttt{0xdc027c5c0d8a6cf88297539240776ebbd64aa094fccff35da6c5ef89b8fc5ae5} & 0.9978 & 135  & 121 \\
$k_{\text{opt}}$    & \texttt{0x9e8a2ae623d20e2830387c6e9bae021ad6355ffd4cf607207bc9eba46c3774da} & 0.9999 & 129 & 127 \\
\hline
\end{tabular}
\label{tab:Entropy_256}
\end{table*}
\subsection{Frequency (Monobit) Test}
The frequency test~\cite{rukhin2001statistical} examines the distribution of binary values within a sequence by measuring the ratio of ones to the total number of bits. It is designed to verify whether the sequence contains an approximately equal number of ones and zeros, as expected in a statistically random bitstream. The underlying assumption is that for a sequence to be truly random, the frequency of ones should be close to one-half.
To evaluate the statistical uniformity of the scalar values, we apply the frequency test to both the randomly selected scalar \(k\) and the optimized scalar \(k_{\text{opt}}\) generated using the proposed DE-based method. The results are summarized in~\ref{tab:Entropy_192}, \ref{tab:Entropy_224}, and~\ref{tab:Entropy_256} focusing on the bit-level distribution of ones and zeros for NIST P-192, NIST P-224, and NIST P-256, respectively.
For the NIST P-192 curve, the random scalar exhibits a distribution of 88 ones and 104 zeros, indicating an imbalance. In contrast, the optimized scalar \(k\) exhibits perfect symmetry, with 96 ones and 96 zeros. Similarly, for NIST P-224, the random scalar shows a skewed distribution of 100 ones and 124 zeros, while the optimized version balances out at 112 ones and 112 zeros. For the NIST P-256 curve, the random scalar again presents a slight bias (135 ones, 121 zeros), whereas the optimized scalar closely approaches uniformity with 129 ones and 127 zeros.
These results demonstrate that the optimized scalars achieve a better bit-level balance compared to their randomly selected scalars. Such uniformity in binary representation is essential in cryptographic contexts, as it minimizes statistical bias and enhances resistance to various analysis and side-channel attacks.
\subsection{Chi-Square Test}
The Chi-square ($\chi^2$) test is commonly used to assess the degree of agreement between the observed and expected frequencies within a categorical dataset. It determines how well the empirical distribution of bit occurrences aligns with the theoretically uniform distribution expected under randomness. The test statistic is computed using the formula~\cite{rukhin2001statistical}:
\begin{equation}
\chi^2 = \sum_{i=1}^{n} \frac{(o_i - e_i)^2}{e_i},
\end{equation}

where $o_i$ represents the observed frequency and $e_i$ denotes the corresponding expected frequency for the $i^{\text{th}}$ category. This statistical measure follows the Chi-square distribution, which is typically right-skewed, and is sensitive to large deviations from expected uniformity. A low $\chi^2$ value indicates strong consistency with the expected distribution, strengthening the randomness of the sequence under test.

We applied the chi-square test to the randomly selected scalars and optimized scalars \( k_{\text{opt}} \) derived using the proposed DE-based method across the NIST P-192, P-224, and P-256 curves. The results summarized 
in~\ref{tab:chi_square_results}, include the chi-square statistic and the corresponding $p$-values for each curve.
For all three curves, the p-values associated with both scalar types were found to be well above the standard rejection threshold of 0.01. This suggests that the null hypothesis that the bit distributions follow a uniform pattern  cannot be rejected. In other words, there is no statistically significant difference between the observed and expected frequencies of bits. Furthermore, the chi-square values for the optimized scalars \( k_{\text{opt}} \) are closer to the ideal region (i.e., lower chi-square statistic and higher $p$-value), indicating a more balanced bit distribution compared to the random scalars. This provides additional evidence that the proposed optimization technique enhances the statistical quality of the scalar, a desirable property for cryptographic robustness.

\begin{table}[htbp]
\centering
\caption{Chi-square~($\chi^2$) test results for random scalar \( k \) and optimized scalar \( k_{\text{opt}} \) across NIST curves.}
\begin{tabular}{|l|l|l|l|}
\hline
\textbf{Curve} & \textbf{Scalar Type} & \textbf{$\chi^2$-value} & \textbf{p-value} \\
\hline
\multirow{2}{*}{NIST P-192} 
& $k$ & 2.8800 & 0.0897 \\
& $k_{\text{opt}}$ & 0.1800 & 0.6714 \\
\hline
\multirow{2}{*}{NIST P-224} 
& $k$ & 4.4138 & 0.0356 \\
& $k_{\text{opt}}$ & 0.2759 & 0.5994 \\
\hline
\multirow{2}{*}{NIST P-256} 
& $k$ & 0.1364 & 0.7119 \\
& $k_{\text{opt}}$ & 0.1364 & 0.7119 \\
\hline
\end{tabular}
\label{tab:chi_square_results}
\end{table}
\subsection{Runs Test}
The runs test~\cite{rukhin2001statistical} focuses on the total runs within a binary sequence. A run is defined as a consecutive series of identical bits (either all ones or all zeros), delimited at both ends by the opposite bit. A run of length \(\ell\) contains exactly \(\ell\) same values bits and is enclosed by bits of the opposing type. The objective of this test is to assess whether the frequency and distribution of these runs, across varying lengths, align with what would be statistically expected in a truly random bit stream. 
The test uses a significance threshold of 1\%. According to the decision rule, if the computed p-value falls below 0.01, the sequence is non-random; otherwise, it is considered random. To ensure meaningful statistical inference, it is recommended that each evaluated sequence contains at least 100 bits.
To assess the bit-level alternation characteristics of the scalar values, we applied the runs test to both the randomly selected scalar \(k\) and the optimized scalar \(k_{\text{opt}}\) across three ECs: NIST P-192, P-224, and P-256. The p-values obtained from the test are summarized in~\ref{tab:run_test_results}.
Across the three NIST curves, both the randomly selected scalars and the optimized scalars produced p-values substantially above the significance threshold of 0.01. This outcome suggests that there is no statistically significant deviation from the expected number of runs in each binary sequence. The results imply that the frequency and distribution of bit transitions (i.e., from 0 to 1 or vice versa) fall within acceptable bounds, reflecting a healthy level of oscillation. This behavior is indicative of sequences with strong randomness properties.
\begin{table}[htbp]
\centering
\caption{Run test p-value comparison for random scalar \( k \) and optimized scalar \( k_{\text{opt}} \) across NIST curves.}
\begin{tabular}{|l|l|l|}
\hline
\textbf{Curve} & \textbf{Scalar Type} & \textbf{p-value} \\
\hline
\multirow{2}{*}{NIST P-192} 
& $k$ & 0.9226 \\
& $k_{\text{opt}}$ & 0.6095 \\
\hline
\multirow{2}{*}{NIST P-224} 
& $k$ & 0.9691 \\
& $k_{\text{opt}}$ & 0.0608 \\
\hline
\multirow{2}{*}{NIST P-256} 
& $k$ & 0.8390 \\
& $k_{\text{opt}}$ & 0.5318 \\
\hline
\end{tabular}
\label{tab:run_test_results}
\end{table}
\subsection{Autocorrelation Test Analysis}
The autocorrelation test assesses the degree of similarity between a binary sequence and a delayed version of itself, quantifying the persistence of patterns across the sequence. This is crucial in cryptographic applications, where lower autocorrelation magnitudes imply reduced predictability and enhanced randomness.
To assess the autocorrelation properties of the scalar values, we computed the autocorrelation coefficients at a set of randomly chosen lags for both the randomly selected scalar \( k \) and the optimized scalar \( k_{\text{opt}} \) generated by the proposed method.
~\ref{tab:autocorrelation_results} presents autocorrelation coefficients computed in ten randomly selected lags. The results show that the optimized scalar \( k_{\text{opt}} \) exhibits significantly lower autocorrelation compared to the randomly selected scalar \( k \) in all lag values. This reduction indicates weaker temporal dependencies and a more uniform distribution of bit transitions, reinforcing the cryptographic suitability of the proposed optimization technique.
\begin{table}[htbp]
\centering
\caption{Autocorrelation coefficients at randomly selected lags where the optimized scalar \( k_{\text{opt}}\) exhibits lower autocorrelation magnitude than the randomly selected scalar \( k \).}
\begin{tabular}{|c|c|c|}
\hline
\textbf{Lag} & \textbf{$k$} & \textbf{$ k_{\text{opt}}$} \\
\hline
2  & -0.1548 & -0.0271 \\
4  & 0.0444  & 0.0043  \\
5  & 0.1642  & 0.0401  \\
16 & -0.0626 & -0.0073 \\
32 & -0.0597 & -0.0093 \\
43 & 0.0317  & -0.0005 \\
45 & -0.1467 & 0.0002  \\
50 & -0.0368 & -0.0044 \\
55 & -0.0107 & 0.0007  \\
60 & 0.0590  & -0.0040 \\
\hline
\end{tabular}
\label{tab:autocorrelation_results}
\end{table}
\section{Conclusion}\label{sec:conclusion}
We proposed an entropy-driven scalar optimization framework for ECC based on the DE algorithm. Unlike traditional random-scalar generation methods, our approach formulates scalar selection as an optimization problem aimed at maximizing bit-level entropy. The resulting scalars exhibit balanced binary distributions, reduced autocorrelation, and improved performance on standard randomness tests, thus strengthening resistance to statistical and side-channel attacks.
To the best of our knowledge, this is the first study to apply evolutionary computation specifically to enhance the entropy of ECC scalars. Our findings reveal that this area has remained largely underexplored, despite the critical role scalar values play in the security of the ECC. By introducing a tunable, deterministic, and repeatable method for scalar selection, we bridge a gap between cryptographic parameter tuning and metaheuristic optimization.
The experimental results validate the effectiveness of the proposed method, showing its potential to improve the cryptographic robustness of ECC without compromising efficiency. This positions our approach as a promising direction for entropy-aware key generation, particularly in security-sensitive applications.
Future work may extend this methodology to other ECC parameters, integrate post-quantum security considerations, or develop hybrid optimization models tailored for resource-constrained environments. Embedding such optimization techniques into practical ECC systems could significantly enhance their resilience and adaptability in evolving threat landscapes.

\section*{Acknowledgement}
During the preparation of this work, the author used a freely available version of ChatGPT to improve the language. After using this tool, the authors reviewed and edited the content as needed and takes responsibility for the content of the published article.

\bibliographystyle{IEEEtran}  
\bibliography{bibfile.bib}
 \end{document}